\documentclass[11pt]{article}
\usepackage{amsfonts}
\usepackage{graphicx}
\usepackage{amsmath}
\usepackage{amssymb}
\usepackage{caption2}
\begin{document}
\bibliographystyle{prsty}
\begin{center}
{\large {\bf \sc{ Analysis of the  light-flavor scalar  and axial-vector  diquark states with QCD sum rules }}} \\[2mm]
Zhi-Gang Wang  \footnote{E-mail:wangzgyiti@yahoo.com.cn. } \\
  Department of Physics, North China Electric Power University, Baoding 071003, P. R.
  China
\end{center}

\begin{abstract}
In this article, we study  the light-flavor scalar and
axial-vector  diquark  states in the vacuum and  in the nuclear matter
 using the QCD sum rules  in an systematic way, and make reasonable predictions for
 their masses in the vacuum and  in the nuclear matter.
\end{abstract}


\section{Introduction}
We can rephrase the scattering amplitude of one-gluon exchange into an antisymmetric  antitriplet  $\overline{\bf{3}}_{ c}$
and an symmetric sextet $\bf{6}_{ c}$ in the color-space,
\begin{eqnarray}
\left(\frac{\lambda^a}{2}\right)_{ki}\left(\frac{\lambda^a}{2}\right)_{lj}&=&-\frac{1}{3}(\delta_{jk}\delta_{il}-\delta_{ik}\delta_{jl})
 +\frac{1}{6}(\delta_{jk}\delta_{il}+\delta_{ik}\delta_{jl})\, ,
\end{eqnarray}
where the $\lambda^a$ is the Gell-Mann matrix element, and the $i,j$
and $k,l$ are the color indexes of the incoming and outgoing quarks respectively.   The attractive interaction
in the  antisymmetric  antitriplet favors the formation of the diquark states in the color
 antitriplet, while  the most stable diquark states maybe exist
in  the color antitriplet $\overline{\bf{3}}_{ c}$, flavor
antitriplet $\overline{\bf{3}}_{ f}$ and spin singlet ${\bf{1}}_s$
channels due to Fermi-Dirac statistics \cite{Color-Spin}.
We can take the diquarks as basic constituents to obtain a new spectroscopy for the mesons
and baryons \cite{Jaffe1977,DiquarkM}, and the diquark states play an important role in many
 phenomenological analysis \cite{diquark-early,ReviewScalar}. For example, we usually take the nonet scalar mesons
  below $1\,\rm{GeV}$   as the tetraquark states consist of the scalar diquark states
$[qq]_{\overline{\bf{3}}_c}$ and $[\bar{q}\bar{q}]_{\bf{3}_c}$ in the relative $S$-wave \cite{ReviewScalar},
and study the octet and decuplet baryons
 as the quark-diquark bound states \cite{Alkofer}.

 The QCD sum rules is a powerful theoretical tool  in
 studying both the in-vacuum and in-medium hadronic properties \cite{SVZ79}, and has been applied extensively
  to study the properties of the in-vacuum hadrons and the in-medium light-flavor hadrons and  charmonium states \cite{NarisonBook,C-parameter,Drukarev1991}.
  In the limit $m_u=m_d\rightarrow 0$, the in-medium nucleon mass $ M_N^*$ can be related with the in-medium quark condensate $\langle\bar{q}q\rangle_{\rho_N}$
  through the simple relation  $M_N^*=-\frac{8\pi^2}{T^2}\langle\bar{q}q\rangle_{\rho_N}$, where the $T^2$ is the Borel parameter. It is
  interesting to study the diquark states in the nuclear matter, as they are basic constituents of the baryons
   and play an important role  in
understanding the strong interactions and the relativistic  heavy ion collisions.
  The in-medium baryon properties will be studies by the CBM (compressed baryonic matter)
 and $\rm{\bar{P}ANDA}$ collaborations unto the charm sector at the upcoming  FAIR (facility for
antiproton and ion research) project at GSI (heavy ion research lab) \cite{CBM,PANDA}.
  The in-vacuum light and heavy scalar
  and axial-vector  diquark states have been studies with the QCD sum rules \cite{Diquark-SR,HuangDiquark,Wang-DQuark}, we extend the previous
  works to study the in-medium mass modifications of the light-flavor  diquark states.

The article is arranged as follows:  we derive the QCD sum rules for
the  light-flavor  scalar and axial-vector diquark states in the vacuum and in the nuclear matter   in Sect.2; in
Sect.3, we present the
 numerical results and discussions; and Sect.4 is reserved for our
conclusions.

\section{ The   scalar and axial-vector  diquark states  with  QCD Sum Rules}
We write down the  two-point correlation functions $\Pi(p)$ and $\Pi_{\mu\nu}(p)$ in the nuclear matter,
\begin{eqnarray}
 \Pi(q)&=&i\int d^4x e^{iqx}\langle \Psi_0|T\{J(x)J^\dagger(0)\}|\Psi_0\rangle \, ,  \nonumber\\
 \Pi_{\mu\nu}(q)&=&i\int d^4x e^{iqx}\langle \Psi_0|T\{J_\mu(x)J_\nu^\dagger(0)\}|\Psi_0\rangle \, ,
\end{eqnarray}
where $J(x)=J^a(x),\eta^a(x)$ and $J_\mu(x)=J_\mu^a(x),\eta_\mu^a(x)$,
\begin{eqnarray}
J^a(x)&=&\epsilon^{abc} u_b^T(x)C\gamma_5 d_c(x)\, , \nonumber\\
\eta^a(x)&=&\epsilon^{abc} q_b^T(x)C\gamma_5 s_c(x)\, ,\nonumber\\
J^{a}_\mu(x)&=&\epsilon^{abc} u_b^T(x)C\gamma_\mu d_c(x)\, , \nonumber\\
\eta^{a}_\mu(x)&=&\epsilon^{abc} q_b^T(x)C\gamma_\mu s_c(x)\, ,
\end{eqnarray}
 the  currents  $J(x)$ and $J_\mu(x)$  interpolate the  scalar and axial-vector diquark states, respectively, the $a$, $b$, $c$
 are the color indexes, the $C$ is the charge conjunction matrix,  and the $ |\Psi_0\rangle$ is the nuclear matter ground state.

In the limit $|\Psi_0\rangle \to |0\rangle$, we  insert  a complete set of intermediate "hadronic" states
 with the same quantum numbers as the current operators $J(x)$ and $J_\mu(x)$ into the correlation functions
 $\Pi(p)$ and $\Pi_{\mu\nu}(p)$  to obtain the
"hadronic" representation \cite{SVZ79}, then isolate  the ground state
contributions from the scalar and axial-vector  diquarks, and obtain  the results:
\begin{eqnarray}
\Pi(q)&=&\frac{f_S^2}{M_S^2-q^2} +\cdots\, , \nonumber\\
\Pi_{\mu\nu}(q)&=&\frac{f_A^2}{M_A^2-q^2}\left(-g_{\mu\nu}+\frac{q_\mu q_\nu}{q^2}\right)+\cdots \, , \nonumber\\
       &=&\Pi(q)\left(-g_{\mu\nu}+\frac{p_\mu p_\nu}{q^2}\right)+\cdots\, ,
\end{eqnarray}
where the pole residues $f_S$ and $f_A$  are defined as
$ \langle 0 | J(0)|S(q)\rangle =f_S$ and $\langle 0 | J_\mu(0)|A(q)\rangle =f_A\epsilon_\mu $,
 the $\epsilon_\mu$ is the polarization vector.

 Here we will take a short digression to discuss the application of the QCD sum rules in studying the diquark states.
  In the QCD sum rules, we perform the operator product expansion at not so deep
 Euclidean space, where the approximation of the correlation functions by  perturbative terms  plus some
  nonperturbative  terms makes sense and the contributions from the condensates (or nonperturbative  terms) are sizeable.
 There are significant differences
between the correlation functions of current operators interpolating the diquarks and conventional handrons,
we can continue the hadronic
correlation functions to the physical region for the conventional hadrons, but not for the diquarks, as the
 diquarks are non-asymptotic states,   there are significant differences  between the diquark states and
  conventional hadrons.
The one-gluon exchange results in strong attractions in the color
antitriplet  channel $\bar{\bf{3}}_c$, the quark-quark system maybe
form quasibound states or loosely bound states (diquark states), which are characterized  by
the correlation length $\mathbb{L}$. At the distance $l>\mathbb{L}$,
the  $\bar{\bf{3}}_c$ diquark state combines with one quark or
one $\bf{3}_c$ antidiquark to form a baryon state or a tetraquark
state, while at the distance $l<\mathbb{L}$, the
 $\bar{\bf{3}}_c$ diquark states dissociate into asymptotic quarks and gluons gradually.
  We can take the diquark state $\mathbb{D}$ as an effective colored hadron and the diquark mass as an
  effective quantity, $M_\mathbb{D}\sim \frac{1}{\mathbb{L}}$,
    the  correlation functions can be continued  to the physical region, where
  the quark-quark correlations exist. The transitions   two-quarks $\leftrightarrow$ diquarks $\leftrightarrow$ hadrons are  not abrupt, the
  typical correlation lengths $\mathbb{L}$  have uncertainties, we have the freedom to choose somewhat larger or smaller diquark masses in model-buildings.  The correlation functions are approximated by a pole term plus a
 perturbative continuum.

We use  the dispersion relation to express the invariant functions  $\Pi(q_0,\vec{q})$ in the following form:
\begin{eqnarray}
\Pi(q_0,\vec{q})&=&{1\over2\pi i}\int_{-\infty}^\infty~d\omega{\Delta\Pi(\omega,\vec{q})\over\omega-q_0} \, ,\nonumber\\
\Delta\Pi(\omega,\vec{q})&=&{\rm{limit}}_{\epsilon\to 0} \left[\Pi(\omega+i\epsilon,\vec{q})-\Pi(\omega-i\epsilon,\vec{q}) \right]\,.
\end{eqnarray}
In the nuclear matter, the corresponding imaginary parts of the spectral densities
can be expressed as
\begin{eqnarray}
\Delta\Pi(\omega,\vec{q})=i\pi\left[F_{+}\delta(q_0-M_+)-F_{-}\delta(q_0+M_-) \right]\,,
\end{eqnarray}
where the $F_{\pm}$ and $M_{\pm}$ are the pole residues and masses, respectively. In the
vacuum limit, $F_{\pm}=\frac{ f_{S}^2 }{M_{S}}, \frac{ f_{A}^2 }{M_{A}}$ and $M_{\pm}=M_S,M_A$.
Thereafter we will use the same notations for the masses and pole residues both in the vacuum and in the nuclear matter
for simplicity.

We carry out the operator product expansion in the finite nuclear density at large spacelike region $q^2\ll 0$,
and express the invariant functions $\Pi(q_0,\vec{q})$ at the level of quark-gluon degrees of freedom
 as \cite{C-parameter,Drukarev1991},
\begin{eqnarray}
\Pi(q_0,\vec{q})&=&\sum_n C_n(q_0,\vec{q})\langle{O}_n\rangle_{\rho_N}\,,
\end{eqnarray}
where the $C_n(q_0,\vec{q})$ are the Wilson coefficients,  the in-medium condensates  $\langle{O}_n\rangle_{\rho_N}=\langle \Psi_0|{O}_n|\Psi_0\rangle =\langle{\cal{O}}\rangle+\rho_N\langle
{\cal{O}}\rangle_N$  at the low nuclear density, the   $\langle{\cal{O}}\rangle$ and $\langle
{\cal{O}}\rangle_N$ denote the vacuum condensates and nuclear matter induced condensates,  respectively,
 then take the limit $u_\mu=(1,0)$, $q_0^2=q^2$,   and obtain the imaginary parts of the QCD spectral densities according to Eq.(5).
 One can consult Refs.\cite{C-parameter,Drukarev1991} for the
technical details in  the operator product expansion. In calculations, we consult  the QCD sum rules
for the light-flavor scalar and axial-vector diquark states in the  vacuum, and
take the analytical expressions of the perturbative terms and the dimension-6 terms from Ref.\cite{Diquark-SR}.

We can match   the phenomenological side with the QCD side of the spectral densities,
and multiply both sides with  the weight function $\omega e^{-\frac{\omega^2}{T^2}}$,  then perform
the integral  $\int_{-\omega_0}^{\omega_0}d\omega$,
\begin{eqnarray}
\int_{-\omega_0}^{\omega_0}d\omega\Delta\Pi(\omega,\vec{q})\omega e^{-\frac{\omega^2}{T^2}}\,,
\end{eqnarray}
where the $\omega_0$ is the threshold parameter,
finally  obtain the following two QCD  sum rules in the nuclear matter:
\begin{eqnarray}
f_{S}^{2}e^{-\frac{M_{S}^2}{T^2}} &=&\frac{3T^4}{4\pi^2}\left[\frac{\alpha_s(T)}{\alpha_s(\mu)}\right]^{\frac{4}{9}}\left\{ \left(1+\frac{17\alpha_s(T)}{6\pi} \right)\left[1-\left(1-\frac{s_0}{T^2}\right)e^{-\frac{s_0}{T^2}} \right]  \right. \nonumber\\
&&\left. -\frac{\alpha_s(T)}{\pi}\int_0^{\frac{s_0}{T2}}dx e^{-x}x {\rm log} x-\frac{2m_s^2}{T^2}\left(1-e^{-\frac{s_0}{T^2}}\right) \right\}-4\langle q^{\dagger}i D_0q\rangle_{\rho_N}-4\langle s^{\dagger}i D_0s\rangle_{\rho_N}\nonumber\\
&&-2m_s\langle \bar{q}q\rangle_{\rho_N}+2m_s\langle \bar{s}s\rangle_{\rho_N}+\frac{m_s\langle \bar{q}g_s\sigma Gq\rangle_{\rho_N}}{T^2}+\frac{8m_s\langle \bar{q}i D_0 i D_0\rangle_{\rho_N}}{T^2}\nonumber\\
&&+\frac{1}{8}\langle \frac{\alpha_sGG}{\pi}\rangle_{\rho_N}+\frac{8\pi\alpha_s\langle\bar{q}q\rangle\langle\bar{s}s\rangle_{\rho_N}}{T^2}
-\frac{16\pi\alpha_s \left[\langle\bar{q}q\rangle_{\rho_N}^2+\langle\bar{s}s\rangle_{\rho_N}^2\right]}{27T^2} \, ,
\end{eqnarray}

\begin{eqnarray}
f_{A}^{2}e^{-\frac{M_{A}^2}{T^2}} &=&\frac{T^4}{2\pi^2}\left[\frac{\alpha_s(T)}{\alpha_s(\mu)}\right]^{\frac{4}{9}}\left\{ \left(1+\frac{\alpha_s(T)}{2\pi} \right)\left[1-\left(1-\frac{s_0}{T^2}\right)e^{-\frac{s_0}{T^2}} \right] -\frac{3m_s^2}{2T^2}\left(1-e^{-\frac{s_0}{T^2}}\right) \right\} \nonumber\\
&& -\frac{8\langle q^{\dagger}i D_0q\rangle_{\rho_N}}{3}-\frac{8\langle s^{\dagger}i D_0s\rangle_{\rho_N}}{3}-2m_s\langle \bar{q}q\rangle_{\rho_N}
+\frac{2m_s\langle\bar{s}s\rangle_{\rho_N}}{3}\nonumber\\
&&+\frac{m_s\langle \bar{q}g_s\sigma Gq\rangle_{\rho_N}}{T^2}+\frac{8m_s\langle \bar{q}i D_0 i D_0\rangle_{\rho_N}}{T^2}-\frac{1}{12}\langle \frac{\alpha_sGG}{\pi}\rangle_{\rho_N}
+\frac{56\pi\alpha_s\langle\bar{q}q\rangle\langle\bar{s}s\rangle_{\rho_N}}{9T^2}\nonumber\\
&&-\frac{4\pi\alpha_s \left[\langle\bar{q}q\rangle_{\rho_N}^2+\langle\bar{s}s\rangle_{\rho_N}^2\right]}{9T^2} \, ,
\end{eqnarray}
where $\alpha_s(T)=\frac{4\pi}{9\log \frac{T^2}{\Lambda^2}}$, $\Lambda =0.375\,\rm{GeV}$, $s_0=\omega_0^2$, $\mu^2=1\,\rm{GeV}^2$.
 We differentiate  Eqs.(9-10) with respect to  $\frac{1}{T^2}$, and obtain two
derived QCD sum rules, then eliminate the
 pole residues $f_{S}$ and $f_A$, and obtain the QCD sum rules for
 the diquark masses. We can replace the mass and condensates of the $s$-quark with the
 corresponding ones of the $q$-quark, and obtain the QCD sum rules for the $ud$ diquark states. In the limit $\rho_N\rightarrow 0$,
 we obtain the corresponding QCD sum rules in the vacuum.
   The renormalization group improvement factor $\left[\frac{\alpha_s(T)}{\alpha_s(\mu)}\right]^{\frac{4}{9}}\approx 1$ at
    the interval $T^2=(0.5-1.1)\,\rm{GeV}^2$, while the terms proportional  to
    $\frac{d}{d\frac{1}{T^2}}\left[\frac{\alpha_s(T)}{\alpha_s(\mu)}\right]^{\frac{4}{9}}$ in the
    derived QCD sum rules have large values, which have significant impacts on the masses of the diquark states both in the
    vacuum and in the nuclear  matter.
    In Fig.1, we plot the values of the $\frac{d}{d\frac{1}{T^2}}\log\left[\frac{\alpha_s(T)}{\alpha_s(\mu)}\right]^{\frac{4}{9}}$ with
    variations of the Borel parameter $T^2$.
\begin{figure}
 \centering
 \includegraphics[totalheight=6cm,width=8cm]{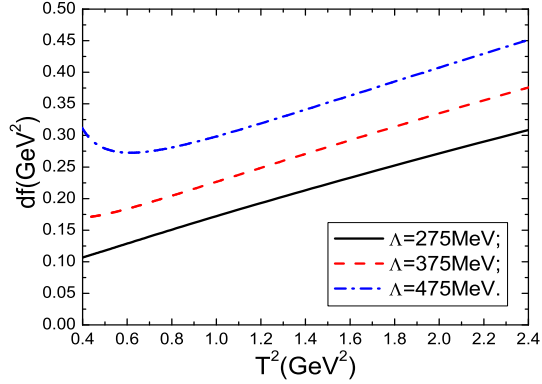}
  \caption{ The values of the $df=\frac{d}{d\frac{1}{T^2}}\log\left[\frac{\alpha_s(T)}{\alpha_s(\mu)}\right]^{\frac{4}{9}}$ with variations of the Borel parameters.  }
\end{figure}

\section{Numerical Results}

In calculations, we have assumed that  the linear density
approximation   is valid at the low nuclear  density. The input parameters are taken as
 $\langle q^\dagger q\rangle_{\rho_N}={3\over2}\rho_N$, $\langle s^\dagger s\rangle_{\rho_N}=0$,
 $\langle \bar{q} q\rangle_{\rho_N}=\langle \bar{q} q\rangle+{\sigma_N\over m_u+m_d}\rho_N $,
 $\langle \bar{s} s\rangle_{\rho_N}=\langle \bar{s} s\rangle+y{\sigma_N\over m_u+m_d}\rho_N $,
 $\langle\frac{\alpha_sGG}{\pi}\rangle_{\rho_N}=\langle\frac{\alpha_sGG}{\pi}\rangle-(0.65\pm0.15)\,{\rm GeV}\rho_N$,
 $\langle q^\dagger iD_0 q\rangle_{\rho_N}=0.18\,{\rm GeV}\rho_N$,
 $\langle s^\dagger iD_0 s\rangle_{\rho_N}=\frac{m_s}{4}\langle \bar{s} s\rangle_{\rho_N}+0.02\,{\rm GeV}\rho_N$,
 $m_u+m_d=12\,\rm{MeV}$, $\sigma_N=(45\pm 10)\,\rm{MeV}$,
$\langle q^{\dagger} iD_0iD_0 q\rangle_{\rho_N}+{1 \over 12}\langle q^{\dagger}g_s\sigma G q\rangle_{\rho_N}=0.031\,{\rm{GeV}}^2\rho_N$,
$\langle s^{\dagger} iD_0iD_0 s\rangle_{\rho_N}+{1 \over 12}\langle s^{\dagger}g_s\sigma G s\rangle_{\rho_N}=y0.031\,{\rm{GeV}}^2\rho_N$,
$\langle\bar{q}g_s\sigma Gq\rangle=m_0^2\langle\bar{q}q\rangle$, $\langle\bar{s}g_s\sigma Gs\rangle=m_0^2\langle\bar{s}s\rangle$,
$\langle \bar{q} iD_0iD_0 q\rangle_{\rho_N}+{1\over 8}\langle\bar{q}g_s\sigma G q\rangle_{\rho_N}=0.3\,{\rm{GeV}}^2\rho_N$,
$\langle \bar{s} iD_0iD_0 s\rangle_{\rho_N}+{1\over 8}\langle\bar{s}g_s\sigma G s\rangle_{\rho_N}=y0.3\,{\rm{GeV}}^2\rho_N$,
$\langle\bar{q}g_s\sigma G q\rangle_{\rho_N}=\langle\bar{q}g_s\sigma G q\rangle+3.0\,{\rm GeV}^2\rho_N$,
$\langle\bar{s}g_s\sigma G s\rangle_{\rho_N}=\langle\bar{s}g_s\sigma G s\rangle+y3.0\,{\rm GeV}^2\rho_N$,
$\langle q^{\dagger}g_s\sigma G q\rangle_{\rho_N}=-0.33\,{\rm GeV}^2\rho_N$, $\langle s^{\dagger}g_s\sigma G s\rangle_{\rho_N}=-y0.33\,{\rm GeV}^2\rho_N$,
  $\langle\bar{q}q\rangle=-(0.23\,\rm{GeV})^3$, $\langle\bar{s}s\rangle=0.8\langle\bar{q}q\rangle$, $m_0^2=0.8\,\rm{GeV}^2$,
$\rho_N=(0.11\,\rm{GeV})^3$, $\langle \bar{q}q\rangle^2_{\rho_N}=f\langle \bar{q}q\rangle_{\rho_N}\times\langle \bar{q}q\rangle_{\rho_N}+(1-f)\langle \bar{q}q\rangle \times \langle \bar{q}q\rangle$, $\langle \bar{s}s\rangle^2_{\rho_N}=f\langle \bar{s}s\rangle_{\rho_N}\times\langle \bar{s}s\rangle_{\rho_N}+(1-f)\langle \bar{s}s\rangle \times \langle \bar{s}s\rangle$, $\langle \bar{q}q\rangle\langle \bar{s}s\rangle _{\rho_N}=f\langle \bar{q}q\rangle_{\rho_N}\times\langle \bar{s}s\rangle_{\rho_N}+(1-f)\langle \bar{q}q\rangle \times \langle \bar{s}s\rangle$, $\langle\frac{\alpha_sGG}{\pi}\rangle=(0.33\,\rm{GeV})^4$,
 $f=0.5\pm 0.5$, $y=0.3$,  $m_s=0.13\,\rm{GeV}$ at the
energy scale  $\mu=1\, \rm{GeV}$ \cite{C-parameter}.

In the conventional QCD sum rules \cite{SVZ79}, there are
two criteria (pole dominance and convergence of the operator product
expansion) for choosing  the Borel parameter $T^2$ and threshold
parameter $s_0$. In this article, we take the pole contributions $R$ as
$(45-80)\%$, just like the ones in our previous works on the heavy, doubly heavy and triply heavy baryon states \cite{Wang-HB},
 the pole contributions $R$ are defined by
\begin{eqnarray}
R&=& \int_{0}^{s_0}dse^{-\frac{s}{T^2} }\rho(s)/\int_{0}^{\infty}dse^{-\frac{s}{T^2}}\rho(s)  \, ,
\end{eqnarray}
where the $\rho(s)$ denotes  the QCD spectral densities, the integral over the $s$ can be carried out analytically, see Eqs.(9-10).
In calculations, we observe that larger threshold parameters lead to larger Borel windows, and choose the possible smallest
Borel windows,  $T^2_{max}-T^2_{min}=0.4\,\rm{GeV}^2$,  to obtain the lowest threshold parameters, and therefore obtain
the possible lowest masses, which correspond to the largest correlation lengths, the revelent values  are
shown explicitly in Table 1. If the same threshold parameters are taken,
 the pole contributions are almost the same in the Borel windows  for the QCD sum rules
 both in the vacuum and in the nuclear matter.
 Thereafter, we will not distinguish the pole contributions
 in the vacuum and in the nuclear matter.
 On the other hand, the main  contributions come from the perturbative terms,
the  operator product expansion are well convergent, the two criteria of the QCD sum rules  are  satisfied.

 Finally we obtain the numerical  values of the masses and pole residues both in the vacuum and in the nuclear matter,
 which are also shown explicitly  in Table 1 and Fig.2. The present predictions
 $\widehat{M}_{ud(0^+)}=0.50\,\rm{GeV}$, $\widehat{M}_{ds(0^+)}=0.64\,\rm{GeV}$ are larger than the values
 $\widehat{M}_{ud(0^+)}=0.40\,\rm{GeV}$, $\widehat{M}_{ds(0^+)}=0.46\,\rm{GeV}$ obtained in Ref.\cite{HuangDiquark}, where
 the derivative $\frac{1}{d\frac{1}{T^2}}$ does not  act on the $\alpha_s(T)$.
From Table 1 and Fig.2, we can see that including the renormalization group improvement factor $\left[\frac{\alpha_s(T)}{\alpha_s(\mu)}\right]^{\frac{4}{9}}$
reduces the diquark masses significantly, $M_{ud(0^+)}-\widehat{M}_{ud(0^+)}=0.14\,\rm{GeV}$,
$M_{qs(0^+)}-\widehat{M}_{qs(0^+)}=0.13\,\rm{GeV}$, $M_{ud(1^+)}-\widehat{M}_{ud(1^+)}=0.11\,\rm{GeV}$,
$M_{qs(1^+)}-\widehat{M}_{qs(1^+)}=0.12\,\rm{GeV}$. Although the factor $\left[\frac{\alpha_s(T)}{\alpha_s(\mu)}\right]^{\frac{4}{9}}\approx 1$,
the terms associate with the  derivative   $\frac{d}{d\frac{1}{T^2}} \left[\frac{\alpha_s(T)}{\alpha_s(\mu)}\right]^{\frac{4}{9}}$ play
 an important role in the derived QCD sum rules. The values of the derivative
  $\frac{d}{d\frac{1}{T^2}}\log\left[\frac{\alpha_s(T)}{\alpha_s(\mu)}\right]^{\frac{4}{9}}$   are rather  large  compared
  with the diquark masses, see Fig.1 and Table 1.

The values of the in-vacuum diquark masses from different theoretical approaches vary in a large range, for examples,
$M_{ud(0^+)}=(0.14-0.74)\,\rm{GeV}$ \cite{BSE1987}, $M_{ud(0^+)}=0.74\,\rm{GeV}$, $M_{qs(0^+)}=0.88\,\rm{GeV}$, $M_{ud(1^+)}=0.95\,\rm{GeV}$,
$M_{qs(1^+)}=1.05\,\rm{GeV}$ \cite{BSE-Burden},
$M_{ud(0^+)}=0.82\,\rm{GeV}$, $M_{qs(0^+)}=1.10\,\rm{GeV}$, $M_{ud(1^+)}=1.02\,\rm{GeV}$,
$M_{qs(1^+)}=1.30\,\rm{GeV}$ \cite{BS-diquark-light},
$M_{ud(0^+)}=0.76\,\rm{GeV}$, $M_{qs(0^+)}=0.98\,\rm{GeV}$ \cite{BSE-Wang} from the Bethe-Salpeter equation with different confining potentials;
$M_{ud(0^+)}=0.710\,\rm{GeV}$, $M_{qs(0^+)}=0.948\,\rm{GeV}$, $M_{ud(1^+)}=0.909\,\rm{GeV}$, $M_{qs(1^+)}=1.069\,\rm{GeV}$ from a relativistic quark
model based on a quasipotential approach in QCD \cite{Ebert2005};
 $M_{ud(0^+)}=(0.694\pm0.022)\,\rm{GeV}$,   $M_{ud(1^+)}-M_{ud(0^+)}=(0.104\pm0.042)\,\rm{GeV}$ from
the lattice QCD \cite{Latt1998};   $M_{ud(0^+)}=(0.42\pm0.03)\,\rm{GeV}$,   $M_{ud(1^+)}=(0.94\pm0.02)\,\rm{GeV}$ \cite{ILM-1994},
$M_{ud(0^+)}=0.5\,\rm{GeV}$ \cite{ILM-2005}  from the random instanton liquid model; $M_{ud(0^+)}= 0.234\,\rm{GeV}$,
$M_{ud(1^+)}=0.824\,\rm{GeV}$ from the  Nambu-Jona-Lasinio Model \cite{Vogl1990};
$M_{ud(0^+)}=0.395\,\rm{GeV}$, $M_{qs(0^+)}=0.590\,\rm{GeV}$ from the constituent diquark model \cite{C-diquark}; etc.
 One should be careful when using them, naively, we expect that they should obey the approximated $SU_f(3)$ symmetry and the
hypersplitting color-spin  interaction maybe  account for the $0^+$ and $1^+$ diquark mass breaking effects.

Lattice QCD calculations   indicate
that the strong attraction in the scalar diquark channels favors the
formation of  good diquarks, the weaker attraction  in the
axial-vector diquark channels maybe form bad diquarks, the energy
gap between the  light-flavor axial-vector and scalar diquarks is about
$\frac{2}{3}$ of the $\Delta$-nucleon mass splitting, $0.2\,\rm{GeV}$ \cite{Latt-SA},
which is also expected from the hypersplitting color-spin  interaction
$\vec{T}_{i}\cdot \vec{T}_{j} \vec{\sigma}_i \cdot\vec{\sigma}_j$ \cite{Color-Spin,ReviewScalar}.
The coupled rainbow Dyson-Schwinger
equation and ladder Bethe-Salpeter equation also indicate such an
energy hierarchy \cite{BS-diquark-light}. In the present work, the central values have the energy gaps
$M_{ud(1^+)}-M_{ud(0^+)}=0.17\,\rm{GeV}$, $M_{qs(1^+)}-M_{qs(0^+)}=0.15\,\rm{GeV}$,
$\widehat{M}_{ud(1^+)}-\widehat{M}_{ud(0^+)}=0.20\,\rm{GeV}$, $\widehat{M}_{qs(1^+)}-\widehat{M}_{qs(0^+)}=0.16\,\rm{GeV}$,
which are consistent with predictions of the lattice QCD and  Bethe-Salpeter equation.
If we neglect the uncertainties, the  $SU_f(3)$ breaking effects for the masses of the scalar and
axial-vector   diquark states are $M_{qs(0^+)}-M_{ud(0^+)}=0.13\,\rm{GeV}$, $M_{qs(1^+)}-M_{ud(1^+)}=0.11\,\rm{GeV}$,
$\widehat{M}_{qs(0^+)}-\widehat{M}_{ud(0^+)}=0.14\,\rm{GeV}$, $\widehat{M}_{qs(1^+)}-\widehat{M}_{ud(1^+)}=0.10\,\rm{GeV}$, respectively,
which are consistent with the naive expectation $m_s-m_q\approx 0.13\,\rm{GeV}$.

From Table 1, we can see that the diquark  states in the nuclear matter have larger masses
  than the corresponding ones in the vacuum, the in-medium effects  lead to the mass-shifts
 $\delta M_{ud(0^+)}=0.04\,\rm{GeV}$,  $\delta M_{qs(0^+)}=0.02\,\rm{GeV}$,
 $\delta \widehat{M}_{ud(0^+)}=0.04\,\rm{GeV}$, $\delta \widehat{M}_{qs(0^+)}=0.01\,\rm{GeV}$,
 $\delta M_{ud(1^+)}=0.06\,\rm{GeV}$, $\delta M_{qs(1^+)}=0.03\,\rm{GeV}$,
 $\delta \widehat{M}_{ud(1^+)}=0.04\,\rm{GeV}$,
 $\delta \widehat{M}_{qs(1^+)}=0.03\,\rm{GeV}$. Although the diquark masses have   uncertainties originate from  the Borel parameters,
 the mass-shifts survive approximately as the uncertainties  are canceled out with each other, see Fig.2,
 the quark-quark correlation   lengths are reduced slightly in the nuclear matter. Compared with the $ud$ diquark states, the $qs$ diquark states have
 much smaller mass-shifts, which   attributes  to the condensates of the $s$-quark obtain much smaller modifications than the corresponding ones of the
  $u$ and $d$-quarks. On the other hand, we can see that the pole residues in the vacuum and in the nuclear matter approximately have the same values,
  as the dominating  contributions  come from the perturbative terms.

\begin{table}
\begin{center}
\begin{tabular}{|c|c|c|c|c|c|c|c|}\hline\hline
              & $T^2/s_0\, (\rm{GeV}^2)$     & pole         & $M(\rm{GeV})$                  & $f(\rm{GeV}^2)$   \\ \hline
            $ud(0^+)$   & $0.5-0.9/1.2$      & $(45-78)\%$  & $0.64\pm0.06\,[0.68\pm0.06]$   & $0.264\pm0.017\,[0.264\pm0.016]$          \\  \hline
            $qs(0^+)$   & $0.6-1.0/1.4$      & $(47-74)\%$  & $0.77\pm0.04\,[0.79\pm0.04]$   & $0.313\pm0.013\,[0.312\pm0.012]$          \\  \hline
            $ud(1^+)$   & $0.6-1.0/1.6$      & $(48-76)\%$  & $0.81\pm0.06\,[0.87\pm0.05]$   & $0.228\pm0.016\,[0.233\pm0.012]$          \\  \hline
            $qs(1^+)$   & $0.7-1.1/1.8$      & $(49-74)\%$  & $0.92\pm0.04\,[0.95\pm0.03]$   & $0.269\pm0.011\,[0.269\pm0.010]$          \\  \hline
  $\widehat{ud(0^+)}$   & $0.5-0.9/1.2$      & $(45-77)\%$  & $0.50\pm0.05\,[0.54\pm0.04]$   & $0.246\pm0.010\,[0.243\pm0.009]$          \\  \hline
  $\widehat{qs(0^+)}$   & $0.6-1.0/1.4$      & $(47-74)\%$  & $0.64\pm0.03\,[0.65\pm0.03]$   & $0.286\pm0.007\,[0.283\pm0.007]$          \\  \hline
  $\widehat{ud(1^+)}$   & $0.6-1.0/1.6$      & $(48-77)\%$  & $0.70\pm0.04\,[0.74\pm0.04]$   & $0.209\pm0.010\,[0.210\pm0.008]$          \\  \hline
  $\widehat{qs(1^+)}$   & $0.7-1.1/1.8$      & $(49-75)\%$  & $0.80\pm0.03\,[0.83\pm0.02]$   & $0.244\pm0.006\,[0.242\pm0.006]$          \\  \hline
                      \hline
\end{tabular}
\end{center}
\caption{ The Borel parameters, threshold parameters, pole contributions, masses and pole residues of  the light-flavor
diquark states, the wide-hat \,$\widehat{}$\, denotes the renormalization group improvement factor $\left[\frac{\alpha_s(T)}{\alpha_s(\mu)}\right]^{\frac{4}{9}}$ is
included, and the bracket denotes the values in the nuclear matter. }
\end{table}

\begin{figure}
 \centering
 \includegraphics[totalheight=5cm,width=6cm]{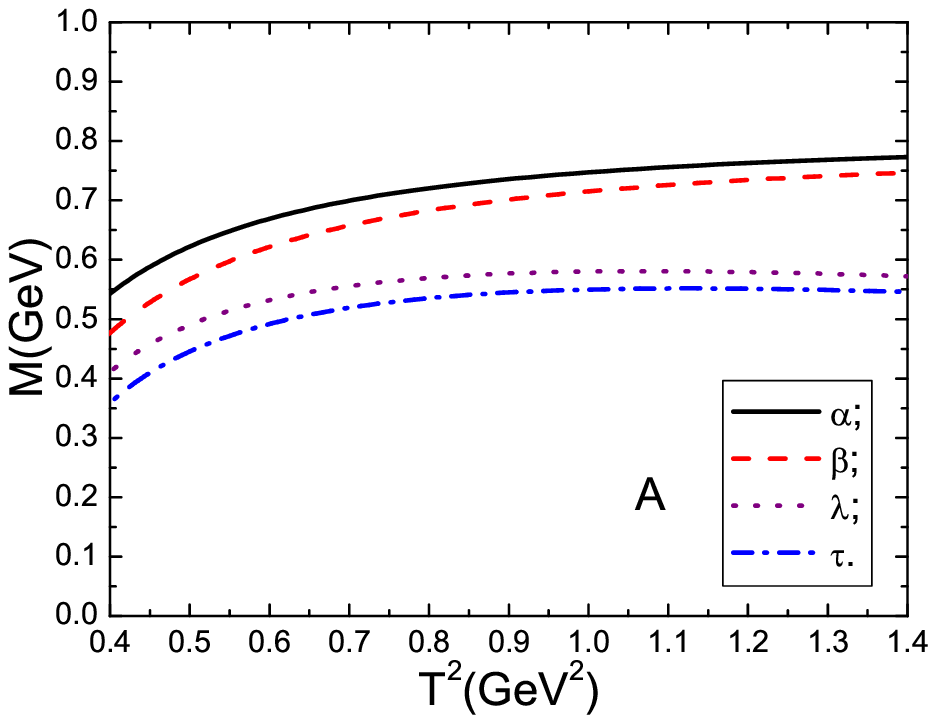}
 \includegraphics[totalheight=5cm,width=6cm]{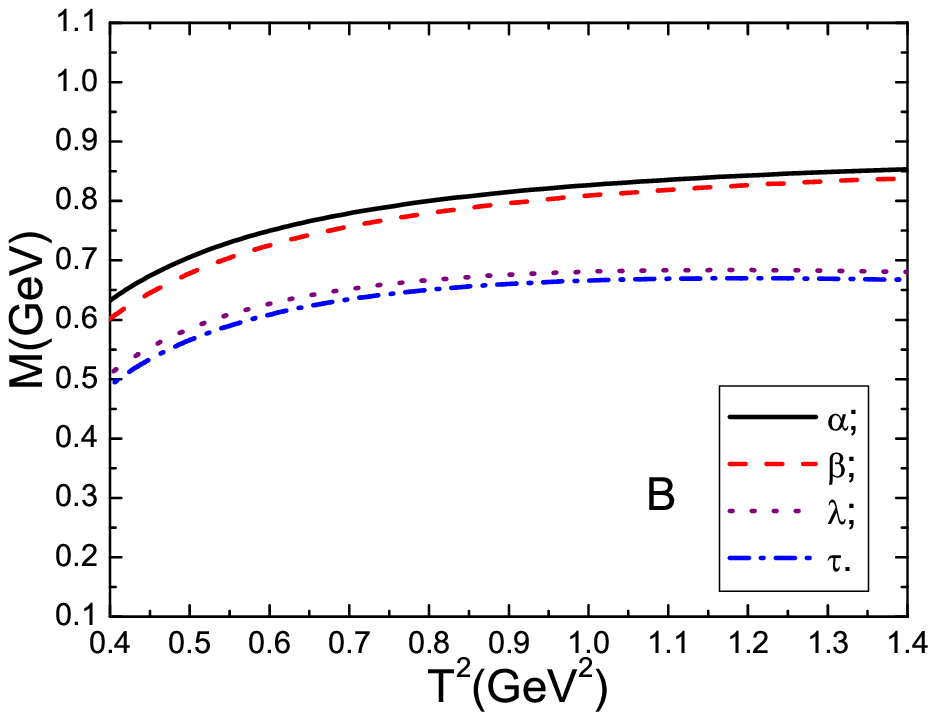}
 \includegraphics[totalheight=5cm,width=6cm]{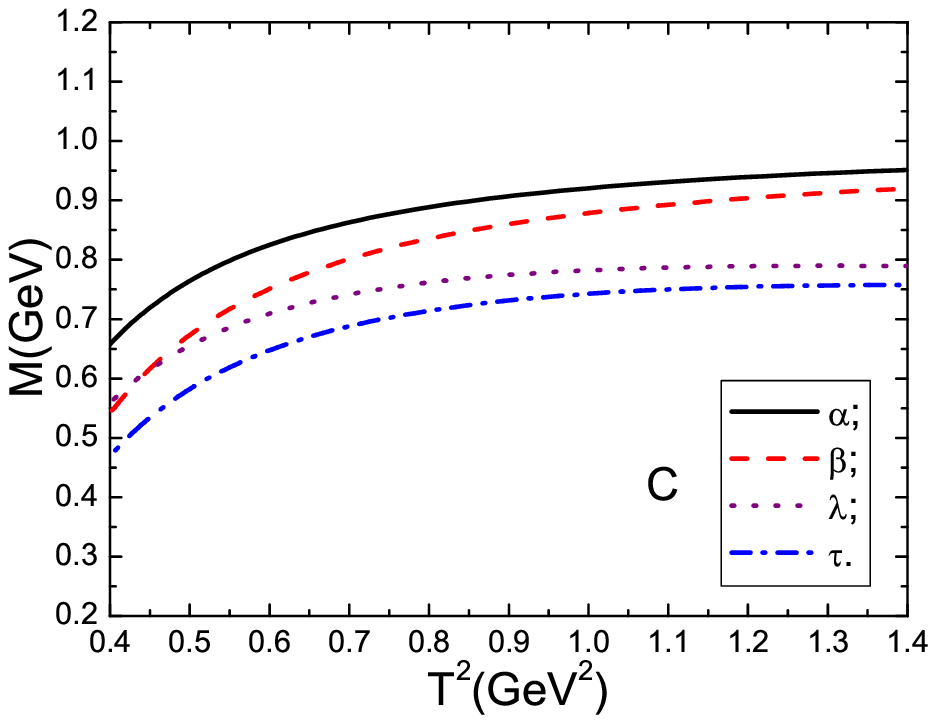}
 \includegraphics[totalheight=5cm,width=6cm]{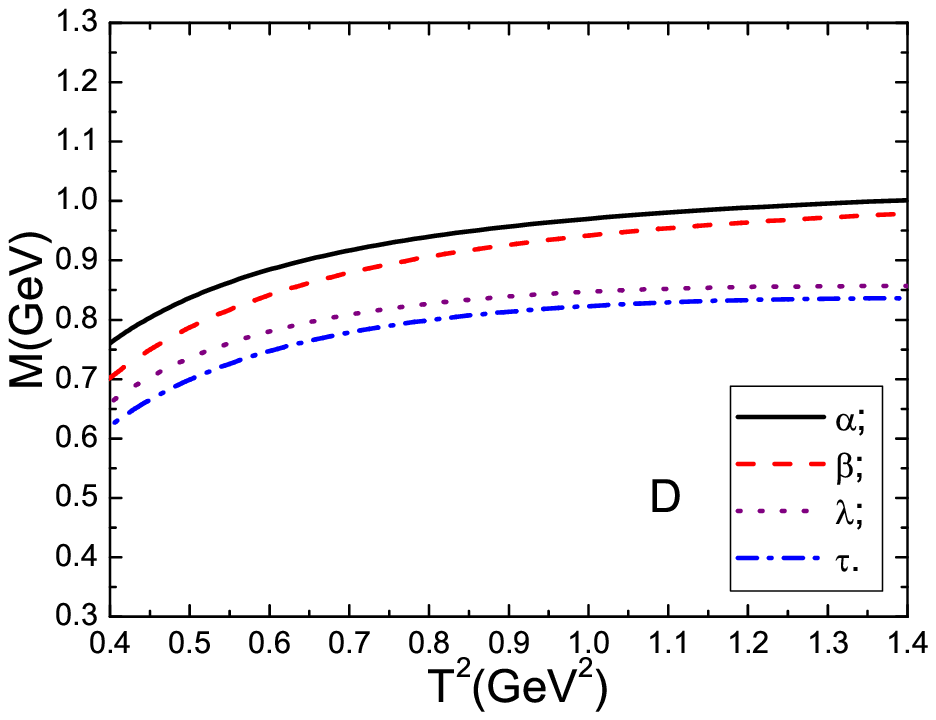}
        \caption{ The masses  $M$ of the light-flavor diquark states with variations of the Borel parameters $T^2$, the $A$, $B$, $C$ and $D$  correspond
      to the $ud(0^+)$, $qs(0^+)$, $ud(1^+)$ and $qs(1^+)$ diquark states respectively. The $\alpha$, $\lambda$ and $\beta$, $\tau$ denote the values from
      the QCD sum rules in the nuclear matter and in the vacuum respectively; while the   $\lambda$, $\tau$ and $\alpha$, $\beta$
      denote the renormalization group improvement factor
      $\left[\frac{\alpha_s(T)}{\alpha_s(\mu)}\right]^{\frac{4}{9}}$ is  included  and not included  respectively.}
\end{figure}

\section{Conclusion}
In this article, we study the light-flavor   scalar and
axial-vector  diquark states in the vacuum and in the nuclear matter
 using the QCD sum rules in an systematic way. The predicted diquark masses in the vacuum
 obey the flavor $SU_f(3)$  symmetry approximately, and the $0^+$ and $1^+$ diquark mass breaking effects are
 consistent with the lattice calculations. The diquark  states in the nuclear matter have larger masses
  than the corresponding ones in the vacuum, the quark-quark correlation   lengths are reduced slightly in the nuclear matter.
  We can take the diquark masses as  basic parameters and perform many phenomenological analysis.

\section*{Acknowledgment}
This  work is supported by National Natural Science Foundation of
China, Grant Number 11075053,  and the
Fundamental Research Funds for the Central Universities.


\begin{thebibliography}{99}

\bibitem{Color-Spin} A. De Rujula, H. Georgi and S. L. Glashow, Phys. Rev. {\bf D12} (1975) 147;
R. L. Jaffe, hep-ph/0001123.

\bibitem{Jaffe1977} R. L. Jaffe, Phys. Rev. {\bf D15} (1977) 267; Phys. Rev. {\bf D15} (1977) 281.

\bibitem{DiquarkM} A. Selem and F. Wilczek, hep-ph/0602128;  T. Friedmann, arXiv:0910.2229.


\bibitem{diquark-early} M. Anselmino, E. Predazzi, S. Ekelin,  S. Fredriksson
and  D. B. Lichtenberg, Rev. Mod. Phys. {\bf 65} (1993) 1199.

\bibitem{ReviewScalar} R. L. Jaffe, Phys. Rept. {\bf 409} (2005) 1.

\bibitem{Alkofer} M. Oettel, G. Hellstern, R. Alkofer, H. Reinhardt, Phys. Rev. {\bf C58} (1998) 2459;
M. Oettel, R. Alkofer,  L. von Smekal, Eur. Phys. J. {\bf A8} (2000) 553.


\bibitem{SVZ79}  M. A. Shifman, A. I. Vainshtein and V. I. Zakharov,
Nucl. Phys. {\bf B147} (1979) 385, 448; L. J. Reinders, H. Rubinstein and S. Yazaki, Phys. Rept. {\bf 127} (1985) 1.

\bibitem{NarisonBook} S. Narison, Camb. Monogr. Part. Phys. Nucl. Phys. Cosmol. {\bf 17} (2002) 1.

\bibitem{C-parameter} T. D. Cohen,  R. J. Furnstahl, D. K. Griegel  and  X. M. Jin, Prog. Part. Nucl. Phys. {\bf 35} (1995)
221; X. M. Jin, T. D. Cohen, R. J. Furnstahl and D. K. Griegel,  Phys. Rev. {\bf C47} (1993) 2882;
X. M. Jin, M. Nielsen, T. D. Cohen,   R. J. Furnstahl and  D. K. Griegel,  Phys. Rev. {\bf C49} (1994) 464.

\bibitem{Drukarev1991} E. G. Drukarev and E. M. Levin,   Prog. Part. Nucl. Phys. {\bf 27} (1991)
77; E. G. Drukarev, M. G. Ryskin and V. A. Sadovnikova, Prog. Part.
Nucl. Phys. {\bf 47} (2001) 73; E. G. Drukarev,   Prog. Part. Nucl. Phys. {\bf 50} (2003) 659.


\bibitem{CBM} B. Friman et al, "The CBM Physics Book: Compressed Baryonic Matter in Laboratory
Experiments", Springer Heidelberg.

\bibitem{PANDA} M. F. M. Lutz et al, arXiv:0903.3905.


\bibitem{Diquark-SR}  H. G. Dosch, M. Jamin and B. Stech, Z. Phys. {\bf C42} (1989)
167;  M. Jamin and M. Neubert, Phys. Lett. {\bf B238} (1990) 387.

\bibitem{HuangDiquark} A. Zhang,  T. Huang and T. G. Steele,  Phys. Rev. {\bf D76} (2007) 036004.

\bibitem{Wang-DQuark} Z. G. Wang,  Eur. Phys. J. {\bf C71} (2011) 1524.


\bibitem{Wang-HB} Z. G. Wang,  Phys. Lett. {\bf B685} (2010) 59;
Z. G. Wang,  Eur. Phys. J. {\bf C68} (2010) 479;
Z. G. Wang,  Eur. Phys. J. {\bf C68} (2010) 459; 
Z. G. Wang, Eur. Phys. J. {\bf A45} (2010) 267;
Z. G. Wang,  Eur. Phys. J. {\bf A47} (2011) 81;
Z. G. Wang,  arXiv:1112.2274.



\bibitem{BSE1987} R. T. Cahill, C. D. Roberts, J. Praschifka,   Phys. Rev. {\bf D36} (1987) 2804.

\bibitem{BSE-Burden} C. J. Burden, L. Qian, C. D. Roberts, P. C. Tandy, M. J. Thomson, Phys. Rev. {\bf C55} (1997) 2649.

\bibitem{BS-diquark-light} P. Maris, Few Body Syst. {\bf 32} (2002) 41.

\bibitem{BSE-Wang} Z. G. Wang, S. L. Wan, W. M. Yang, Commun. Theor. Phys. {\bf 47} (2007) 287.

\bibitem{Ebert2005} D. Ebert, R. N. Faustov, V. O. Galkin, Phys. Rev.{\bf  D72} (2005) 034026.

\bibitem{Latt1998} M. Hess, F. Karsch, E. Laermann, I. Wetzorke, Phys. Rev. {\bf D58} (1998) 111502.

\bibitem{ILM-1994} T. Schaefer, E. V. Shuryak, J. Verbaarschot,  Nucl. Phys. {\bf B412} (1994) 143.

\bibitem{ILM-2005} M. Cristoforetti, P. Faccioli, G. Ripka, M. Traini, Phys. Rev. {\bf D71} (2005) 114010.

\bibitem{Vogl1990} U. Vogl,   Z. Phys. {\bf A337} (1990) 191.

\bibitem{C-diquark} L. Maiani, A. D. Polosa, F. Piccinini, V. Riquer,  Phys. Rev. Lett. {\bf 93} (2004) 212002;
L. Maiani, F. Piccinini, A. D. Polosa, V. Riquer, Phys. Rev. {\bf D71} (2005) 014028.

\bibitem{Latt-SA} C. Alexandrou,  P. de Forcrand and  B. Lucini,  Phys. Rev. Lett. {\bf 97} (2006) 222002.


\end{thebibliography}
\end{document}